\documentclass[structabstract]{aa}  
\usepackage{graphicx}
\usepackage{txfonts}
\usepackage[authoryear]{natbib}
\usepackage{longtable}
\usepackage{rotating}
\usepackage{lscape}
\bibpunct[; ]{(}{)}{;}{a}{}{,}
\begin{document}

\title{Discovery of New Companions to High Proper Motion Stars\\
from the VVV Survey.
\thanks{The data for this work were obtained via observing 
programmes 179.B-2002, 087.D-0490, and 089.D-0462.}}

\author{
Valentin D. Ivanov\inst{1}
\and
Dante Minniti\inst{2,3,4,5}
\and
Maren Hempel\inst{2,5}
\and
Radostin Kurtev\inst{6}
\and
Ignacio Toledo\inst{2,7}
\and
Roberto K. Saito\inst{2,5,6,8}
\and
Javier Alonso-Garc\'ia\inst{2,5}
\and
Juan Carlos Beam\'in\inst{2,5,9}
\and
Jura Borissova\inst{5,6}
\and
M\'arcio Catelan\inst{2,5}
\and
Andr\'e-Nicolas Chen\'e\inst{6,10,11}
\and
Jim Emerson\inst{12}
\and
\'Oscar A. Gonz\'alez\inst{1}
\and
Phillip W. Lucas\inst{13}
\and
Eduardo L. Mart\'in\inst{14}
\and
Marina Rejkuba\inst{15}
\and
Mariusz Gromadzki\inst{6}
}

\offprints{V. Ivanov, \email{vivanov@eso.org}}

\institute{
European Southern Observatory, Ave. Alonso de C\'ordova 3107, 
Vitacura, Santiago, Chile
\and 
Instituto de Astrof\'isica, Facultad de F\'isica, Pontificia 
Universidad Cat\'olica de Chile, Casilla 306, Santiago 22, Chile
\and
Departamento de Ciencia F\'isicas, Universidad Andr\'es Bello, 
Santiago, Chile
\and
Vatican Observatory, V00120 Vatican City State, Italy
\and
The Milky Way Millennium Nucleus, Av. Vicu\~{n}a Mackenna 4860, 
782-0436, Macul, Santiago, Chile
\and
Departamento de F\'isica y Astronom\'ia, Facultad de Ciencias, 
Universidad de Valpara\'iso, Av. Gran Bretana 1111, Playa Ancha, 
Casilla 5030, Valpara\'iso, Chile
\and
ALMA Santiago Central Offices, Alonso de C\'ordova 3107, Vitacura,
Casilla 763 0355, Santiago, Chile
\and
Universidade Federal de Sergipe, Departamento de F\'isica, Av.
Marechal Rondon s/n, 49100-000, S\~ao Crist\'ov\~ao, SE, Brazil
\and
Instituto de Astrof\'isica, Facultad de F\'isica, Pontiﬁcia 
Universidad Cat\'olica de Chile, Avda. Vicu\~na Mackenna 4860, 
782-0436 Macul, Santiago, Chile 
\and
Departamento de Astronom\'ia, Universidad de Concepci\'on, 
B\'io-B\'io 160-C, Concepci´on, Chile
\and
Gemini Observatory, Northern Operations Center, 670 North A'ohoku 
Place, Hilo, HI 96720, USA
\and
Astronomy Unit, School of Physics \& Astronomy, Queen Mary 
University of London, Mile End Road, London, E1 4NS, UK
\and
Centre for Astrophysics Research, University of Hertfordshire, 
College Lane, Hatfield AL10 9AB, UK
\and
Centro de Astrobiolog\'ia (INTA-CSIC), Carretera de Ajalvir, 
km 4, E-28850 Torrej\'on de Ardoz, Madrid, Spain
\and
European Southern Observatory, Karl-Schwarszchild-Str. 2, 
D-85748 Garching bei Muenchen, Germany
}

\date{Received 2 November 1002 / Accepted 7 January 3003}

\abstract 
{The severe crowding in the direction of the inner
Milky Way suggests that the census of stars within a few tens of
parsecs in that direction may not be complete.}
{We search for new nearby objects companions of known high 
proper motion (HPM) stars located towards the densest 
regions of the Southern Milky Way where the background 
contamination presented a major problem to previous works.}
{The common proper motion (PM) method was used--we inspected 
the area around 167 known HPM ($\geq$200\,mas\,yr$^{-1}$) 
stars: 67 in the disk and 100 in the bulge. Multi-epoch 
images were provided by the Two Micron All Sky Survey (2MASS) 
and the VISTA Variables in Via Lactea (VVV). The VVV is a new 
on-going $ZYJHK_S$ plus multi-epoch $K_S$ survey of 
$\sim$562\,deg$^2$ of Milky Way's bulge and inner Southern 
disk.}
{Seven new co-moving companions were discovered around known
HPM stars (L\,149-77, LHS\,2881, L\,200-41, LHS\,3188, 
LP\,487-4, LHS\,5333, and LP\,922-16); six known co-moving 
pairs were recovered (LTT\,5140\,A + LTT\,5140\,B, L\,412-3 + 
L\,412$-$4, LP\,920$-$25 + LP\,920$-$26, LTT\,6990\,A + 
LTT\,6990\,B, M\,124.22158.2900 + M\,124.22158.2910, and 
GJ\,2136\,A + GJ\,2136\,B);
a pair of stars that was thought to be co-moving was found 
to have different proper motions (LTT\,7318, LTT\,7319);
published HPMs of eight stars were not confirmed (C*\,1925, 
C*\,1930, C*\,1936, CD$-$60\,4613, LP\,866$-$17,
OGLE\,BUL$-$SC20\,625107, OGLE\,BUL$-$SC21\,298351, and 
OGLE\,BUL$-$SC32\,388121);
last but not least, spectral types ranging from G8V to M5V were 
derived from new infrared spectroscopy for seventeen stars, 
members of the co-moving pairs.}
{The seven newly discovered stars constitute $\sim$4\% of
the nearby HPM star list but this is not a firm limit on the
HPM star incompleteness because our starting point--the HPM 
list assembled from the literature--is incomplete itself, 
missing many nearby HPM M and L type objects, and it is 
contaminated with non-HPM stars. We have demonstrated, that 
the superior sub-arcsec spatial resolution, with respect to 
previous surveys, allows the VVV to examine further the 
binary nature nature of known HPM stars. The $\geq$5\,yr 
span of VVV will provide sufficient baseline for finding 
new HPM stars from VVV data alone.
}

\authorrunning{V. D. Ivanov et al.}
\titlerunning{New Companions to High Proper Motion Stars from 
the VVV Survey}
\maketitle

\section{Introduction}

The direction toward the inner Milky Way presents a formidable 
challenge for proper motions (PM) studies because of the crowding 
and 
confusion \citep[for previous attempts see ][]{2002AJ....124.1190L,
2008AJ....135.2177L}. {\it VISTA Variables in the Via Lactea} (VVV) 
is a new ESO Public survey \citep{2010NewA...15..433M,
2012A&A...537A.107S} that may help to alleviate these problems. The 
VVV is carried out with the {\it Visible and Infrared Survey 
Telescope for Astronomy} \citep[VISTA; ][]{Dalton2006,Emerson2006}
at Paranal Observatory, and will 
obtain $ZYJHK_S$ coverage and multi-epoch (up to 100 for some 
pointings) $K_S$ observations of $\sim$562\,deg$^2$ in the Milky 
Way's bulge and inner disk, at sub-arcsec seeing. After two years 
of operation, we already demonstrated, that VVV is producing new, 
interesting results: discovery of new star clusters 
\citep{2011A&A...527A..81M,2011A&A...532A.131B,2011A&A...535A..33M},
investigation of the structure and stellar populations content of 
the Milky Way \citep{2011ApJ...733L..43M,2011A&A...534L..14G,
2011A&A...534A...3G,2012A&A...543A..13G,2012A&A...544A.147S}, study 
of variable stars and transients \citep{2011arXiv1105.1119C,
2012ATel.4426....1S}, and others. 
One of the main goals is to obtain 3-dimensional tomographic map of 
the Milky Way bulge based on Red Clump giants, RR\,Lyr and Cepheid 
variables. However, some corollary science objectives are also 
considered, 
including a PM study, taking advantage of the projected $\geq$5\,yr 
survey duration. Early PM science with VVV is possible if it is used 
as a second epoch to a previous infrared survey, $e.g.$ the Two 
Micron All Sky Survey \citep[2MASS;][]{2006AJ....131.1163S}. The 
2MASS observations provide $\geq$10\,yr baseline.

The VVV footprint on the sky is relatively small:
$\sim$20\,deg\,$\times$ 15\,deg centered on the bulge, and 
$\sim$55\,deg\,$\times$\,4.5\,deg along the adjacent Southern disk, 
or just above 1\% of the total sky, but it encompasses the regions 
with the highest stellar surface density in the Galaxy. This study 
is based on the available multi-filter imaging that was taken during 
the first VVV observing season, covering $\sim$500\,deg$^2$. The 
typical image quality is 0.8-1.0\,arcsec, and the pixel scale is 
$\sim$0.34\,arcsec\,pix$^{-1}$, which compare favorably to other 
Galactic surveys. The final VVV data products will be $ZYJHK_S$ 
atlas and a catalog of $\sim$10$^9$ sources, $\sim$10$^6$ of which 
are variables\footnote{For further details see the VVV web page at: 
{\it http://vvvsurvey.org}}.

We embarked on a project to improve the Solar neighborhood census 
by searching for common PM companions to known nearby HPM stars. Our 
effort has the potential to improve the local stellar multiplicity 
fraction estimate--a key constraint to star formation theories, 
with implications for the stellar population modeling of unresolved 
stellar systems. We were driven by the argument that relatively 
bright new solar neighborhood stars could be found only in a 
survey covering the densest regions of the Milky Way, like VVV,
because such stars far away from the Galactic plane were easy to
discover with the previous generation of surveys.
We build upon the success of the RECONS PM and 
parallax measurements project \citep[][]{2007AJ....133.2898F,
2004AJ....128.2460H,2009AJ....137.3800J,2009AJ....137.4547S}, 
the work of L\'eipne and collaborators \citep{2002AJ....123.3434L,
2007AJ....133..889L}, Raghavan et al. \citep{2010ApJS..190....1R},
Faherty et al. \cite{2010AJ....139..176F}, Allen et al. 
\citep{2012AJ....144...62A}, and others, with the advantage that 
VVV has better spatial resolution and higher sensitivity to low 
mass red objects, with respect to the previous surveys. 

The paper is organized as follows: the next section describes the 
sample, Sec.\,\ref{sec:method} summarizes the search method, the
follow up spectroscopy is reported in Sec.\,\ref{sec:spectra}, and 
Sec.\,\ref{sec:results} gives the results.

\section{Sample Selection}\label{sec:sample}

The sample was selected with SIMBAD, and it includes all stars 
with PM$\geq$200\,mas\,yr$^{-1}$, implying a total movement of 
$\geq$2\,arcsec ($\geq$6 VIRCAM pixels) over the $\sim$10-year 
interval separating 2MASS\,and VVV. Our previous experience shows 
that movements of this magnitude are easily detected. Some of the 
stars are saturated on the VVV images, but this still allows to 
search for fainter co-moving companions because the motions of 
the ``doughnuts'' with burned out cores are still discernible.

The VVV survey plan for the first year envisioned separate visits 
of each point of the survey area for $ZY$ and $JHK_S$ observations, 
and up to six visits in $K_S$, on separate nights, for variability 
studies. At the time we carried this study, 144 out of 152 disk 
tiles and 188 out of 196 bulge tiles were completed, covering 
$\sim$216 and $\sim$282\,deg$^2$, respectively.

We arrived at a target list of 167 objects: 67 in the disk and 100 
in the bulge, for which the VVV can provide a new epoch of 
observations (Table\,\ref{table:targets}). The stars come from the 
catalogs of 
LHS (Luyten 1979a), 
LTT \citep{1957QB6.L98........,1961QB811.L853......,1962QB6.L985.......},
MACHO\,\citep[Alcock et al.][]{2001ApJ...562..337A}, 
NLTT \citep{1979nltt.book.....L,1979nlccs.book.....L,1980nltt.bookQ....L},
OGLE\,\citep{2004MNRAS.348.1439S}, and 
number of other works: 
Finch et al. (2007)  
L\'epine \citep{2005AJ....130.1247L}, 
Rattenbury \& Mao \citep{2008MNRAS.385..905R},
Subasavage et al. \citep{2005AJ....130.1658S},
Terzan et al. \citep{1980CRASB.290..321T}.

The selection is dominated by nearby dwarfs (SIMBAD lists: 8 F, 19 
G, 24 K, and 22 M-types). The distances to the few stars with 
parallaxes range from $\sim$1.3 to $\sim$136\,pc, with a median 
of $\sim$44\,pc. 

The sample is heterogeneous, subjected to different biases, and the 
PMs from different catalogs have different error budgets, so our work 
cannot present a basis for strict statistical studies of the Solar 
neighborhood. This question will be addressed after the completion 
of VVV, when it will generate a long baseline coverage with 
self-consistent data.

\section{Analysis}\label{sec:method}

\begin{figure*}[t!]
\includegraphics[height=6.07cm,width=6.07cm]{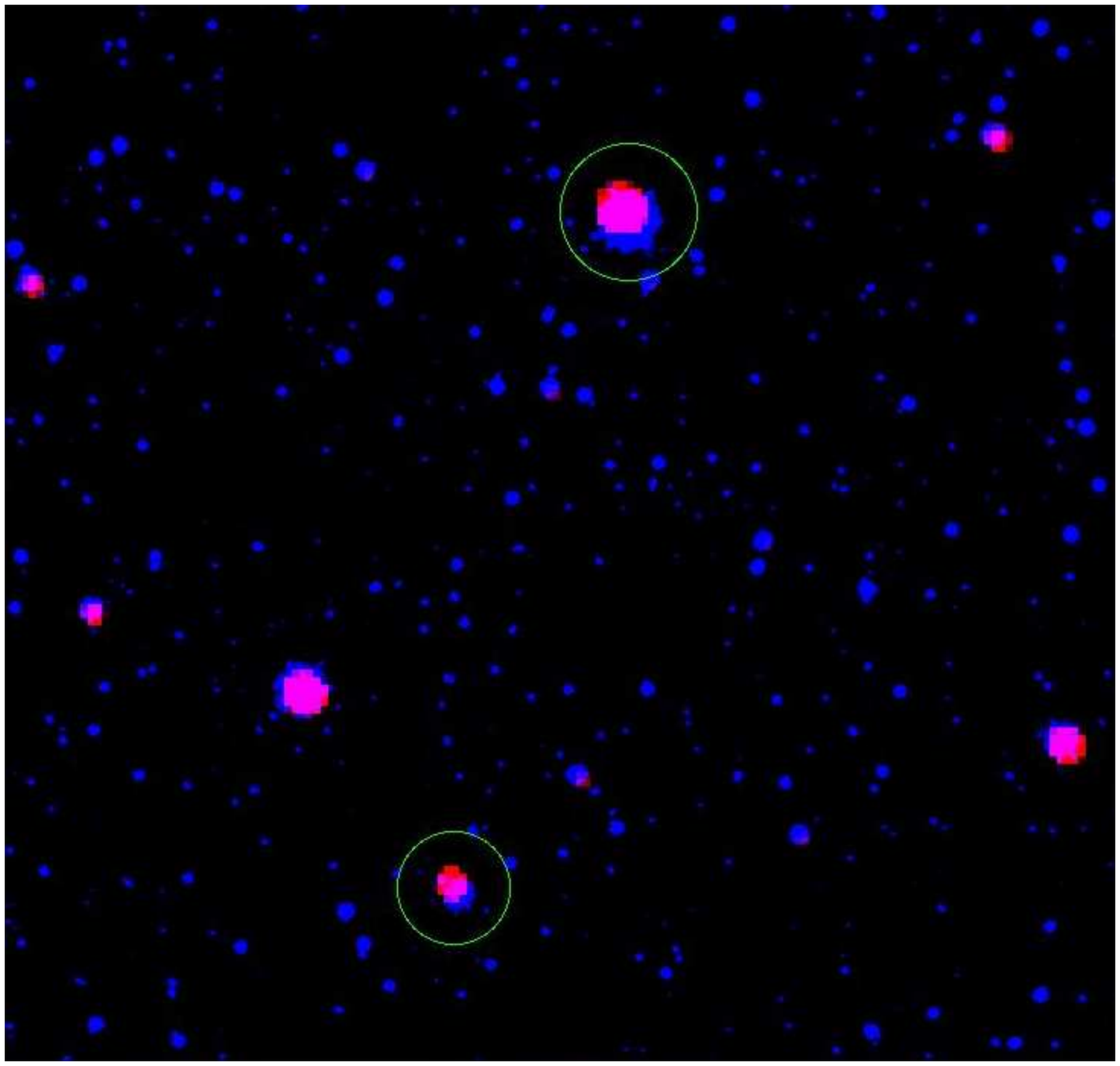}
\includegraphics[height=6.07cm,width=6.07cm]{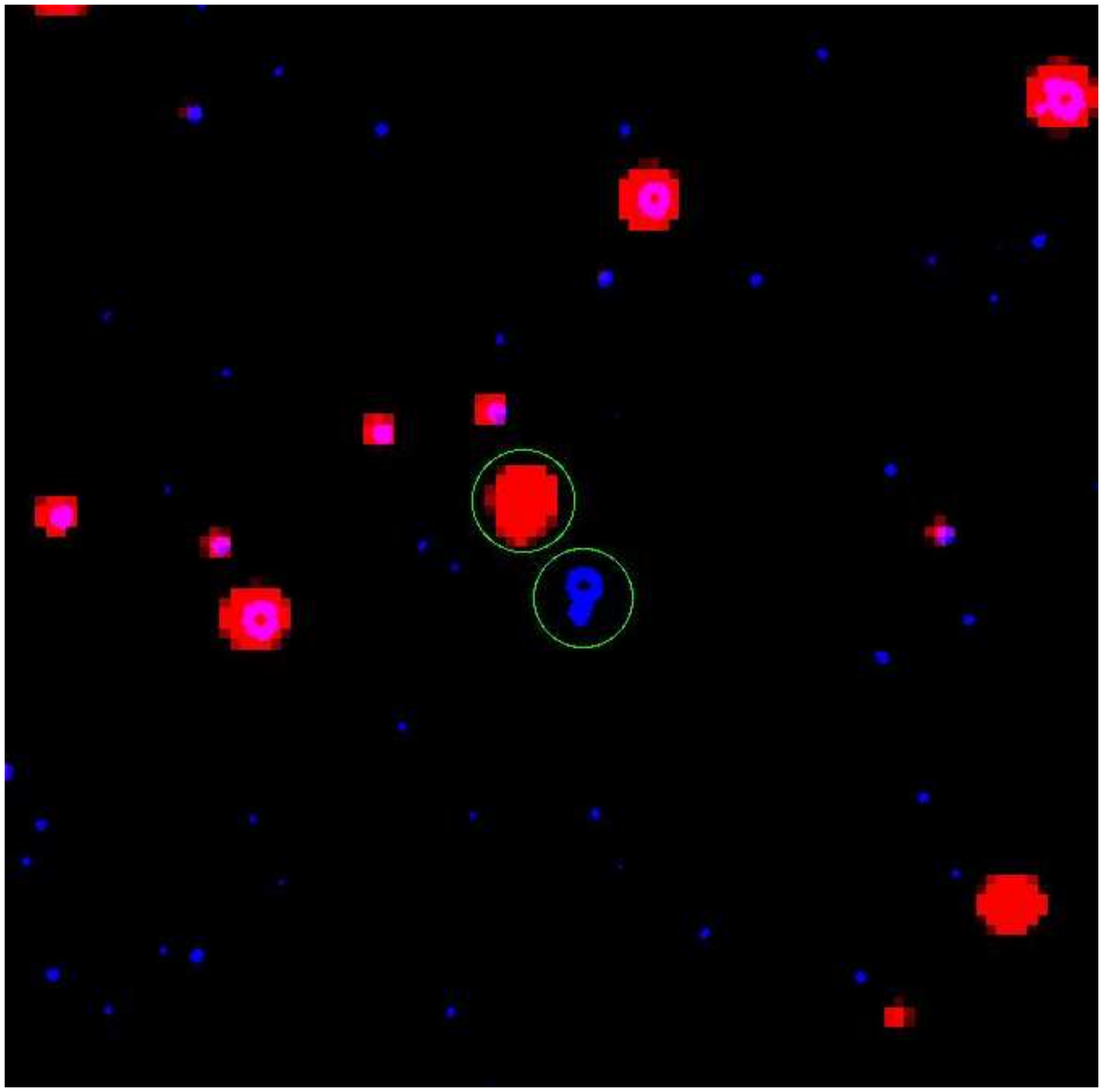}
\includegraphics[height=6.07cm,width=6.07cm]{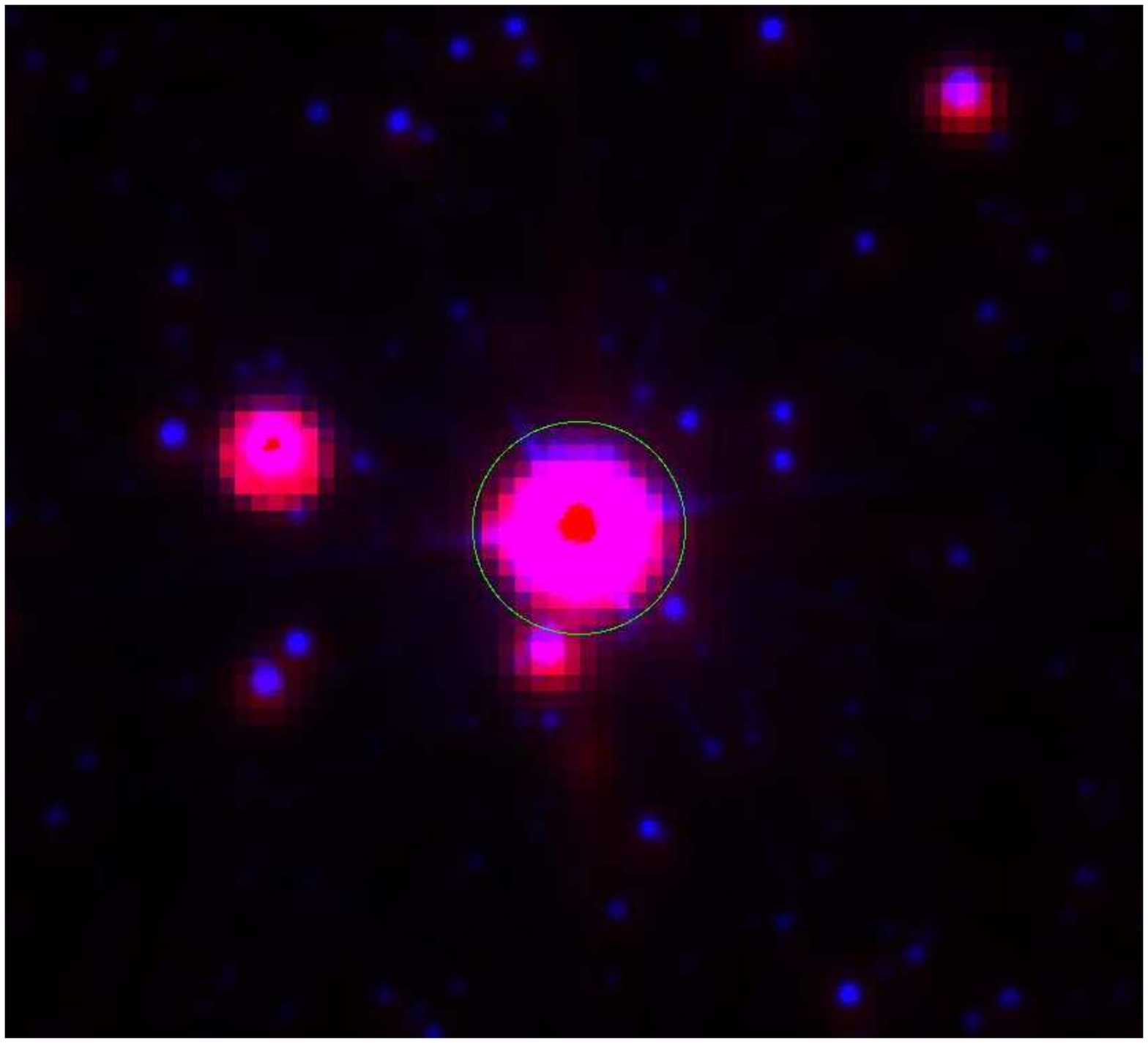}
\caption{Examples for the co-moving companion search. North is up 
and East is left. The red images are 2MASS $K_S$, and the blue 
images are VVV $K_S$. The objects of interest are circled. From 
left to right: LP\,922-16\,A and B ($\sim$2$\times$2\,arcmin),
LHS\,3188\,A and B ($\sim$1.5$\times$1.5\,arcmin), and C*\,1936 
($\sim$1$\times$1\,arcmin).}\label{fig:seach_example}
\end{figure*}

We visually searched around known HPM stars on false-3-color 
images, generated combining the reddest available POSSII band, 
and $J$-bands from 2MASS\,and VVV (Fig.\,\ref{fig:seach_example}). 
Three field sizes were used, to provide different levels of 
``zoom'' into the vicinity of program stars: 0.9, 1.8, and 
3.6\,arcmin, centered on each candidate. Inspecting larger 
images was found impractical. 

For each object in our sample we performed the following steps:
first, we identified the known HPM star from its coordinates 
and the apparent change of position between the epochs of the 
surveys used to create the false-3-color images. The large PM 
made these identifications unambiguous. Second, we selected 
candidate companions looking by eye for other stars in the 
field with similar apparent motion as the known HPM star. Some 
candidates were discarded later, after we calculated their PMs 
(see below) and found them inconsistent with the PMs of the 
known HPM star. In a few cases the candidate companions were 
not fully resolved on the older images. Then, we used the 
non-circular PSF of older surveys as an argument supporting the 
companionship (Fig.\,1, middle). Finally, we inspected the 
selected candidates on 3-color images built from VVV $JHK_S$ 
data to minimize missidentification, $e.g.$ rising from extreme 
colors or artifacts. The subjective nature of this procedure,
together with the varying point spread function (PSF) are 
difficult to quantify, making our results unsuitable for a 
rigorous statistical constrain on the completeness of HPM stars.

The astrometric calibration of VVV data is based on hundreds
of 2MASS stars that fall onto each tile \citep{2004SPIE.5493..411I,
2010NewA...15..433M}. This procedure removes the systematic bulk 
motion of the ``unmoving'' background stars between the VVV and the 
2MASS epochs. Therefore, we directly compared 2MASS and VVV 
coordinates, to measure PMs. This makes our PMs relative in nature, 
because the 2MASS reference stars that were used to derive the 
astrometric solution for the VVV have some average common motion 
than remains unaccounted for.

We measured PMs only for the new co-moving HPM candidates, for 
their hosts from the known HPM star list, and for the stars from 
the HPM list that appeared to move with much slower PM than the 
one given in the literature (Tables\,\ref{table:multi-epoch_coo}, 
and 3).
We calculated the 
stellar positions as unweighted centroids. The cores of the bright 
stars ($K_S$$\leq$12\,mag) are saturated, and to investigate the 
effect of the saturation we set to zero the central pixels that 
are above 60\% of the saturation limit, for 50 stars below the 
saturation limit. The result was much stronger than the typical 
saturation effect for the stars in our sample. The differences of 
the coordinates with and without ``saturation'' was 
0.03$\pm$0.03\,arcsec, $e.g.$ the wings of the images are 
sufficient to measure the stellar positions accurately.

The final PMs are simple arithmetic averages  of the PMs determined
between 2MASS and various VVV observations, and the error are 
the r.m.s. of the measurements, if more than three are available 
(Tables\,\ref{table:companions} and \ref{table:not_high_pm}). 
Adding older photographic epochs usually worsens the fit because 
of crowding and contamination. The 2MASS sets the faint magnitude 
limit of our new HPM candidates, and the minimum primary--companions 
separation: $J$$\leq$16\,mag and d$\geq$1.5-1.7\,arcsec, 
respectively \citep{2006AJ....131.1163S}. The maximum separation 
was determined by the size of the cut outs. 

After inspecting the 167 objects in our sample (67 in the disk and 
100 in the bulge), we found:

(1) seven new co-moving companions to bright 
($J$$\leq$16\,mag) HPM stars with PM$\leq$200\,mas\,yr$^{-1}$:
L\,149-77, LHS\,2881, L\,200-41, LHS\,3188, LP\,487-4, 
LHS\,5333, and LP\,922-16. Particularly notable is the discovery 
of a low-mass M5V companion to LHS\,3188, at $\sim$21\,pc from 
the Sun;

(2) six known co-moving binaries were recovered: 
LTT\,5140\,A + LTT\,5140\,B, 
L\,412$-$3 + L\,412$-$4, 
LTT\,6990\,A + LTT\,6990\,B, 
GJ\,2136\,A + GJ\,2136\,B,  
LP\,920$-$25 + LP\,920$-$26, and
MACHO\,124.22158.2900 + MACHO\,124.22158.2910;

(3) LTT\,7318 and LTT\,7319 that were considered co-moving 
stars, appeared not to be;

(4) we measured the PMs of all co-moving pairs of HPM stars in 
our sample (Table\,\ref{table:companions}), and of the stars with
previously overestimated PMs (Table\,\ref{table:not_high_pm});

(5) the spectral types of seventeen members of the co-moving 
pairs were determined from new near-infrared spectroscopy 
(Table\,\ref{table:spec_obs_log}). They range from G8V to M5V;

(6) HPMs of eight stars (C*\,1925, C*\,1930, C*\,1936,
CD$-$60\,4613, LP\,866$-$17, OGLE\,BUL$-$SC20\,625107, 
OGLE BUL$-$SC21\,298351, and OGLE\,BUL$-$SC32\,388121) reported 
in at least some previous works appear to have been grossly 
overestimated.

\section{Follow-up Spectroscopic Observations}\label{sec:spectra}

Near-infrared spectra of co-moving pairs were obtained to determine 
their spectral types at the ESO NTT with SofI 
\citep[Son of ISAAC;][]{1998Msngr..91....9M} in two low-resolution 
modes, with blue ($\lambda$=0.95$-$1.64\,$\mu$m) and red 
($\lambda$=1.53$-$2.52\,$\mu$m) grisms to cover the entire near-infrared
spectral range. The slit was 1\,arcsec wide during the Apr 2011 
run, and 0.6\,arcsec during the May 2012 run, delivering an average 
resolution of R$\sim$600 and $\sim$1000, respectively. It was 
aligned along the axis connecting the two candidate companions, 
except if their apparent magnitudes were too different--in which 
case they were observed separately. Typically, four (six in the 
case of the relatively faint MACHO\,124.22158.2900--MACHO\,124.22158.2910 
binary candidate) images were obtained, into a two-nodding ABBA or 
ABBAAB sequences, with nodding of 30-60\,arcsec. Each image 
constituted 48--1050\,sec of integration, averaged over 3--12 
individual detector integrations to ensure the peak values are well 
below the non-linearity limit of the detector 
(Table\,\ref{table:spec_obs_log}). The atmospheric 
conditions varied during the observations but most often they were 
mediocre, with a seeing above 1.5\,arcsec, thin to thick 
cirrus--because these targets were poor weather fillers which 
accounts for somewhat longer than usual integration times.

\begin{figure}
\resizebox{\hsize}{!}{\includegraphics{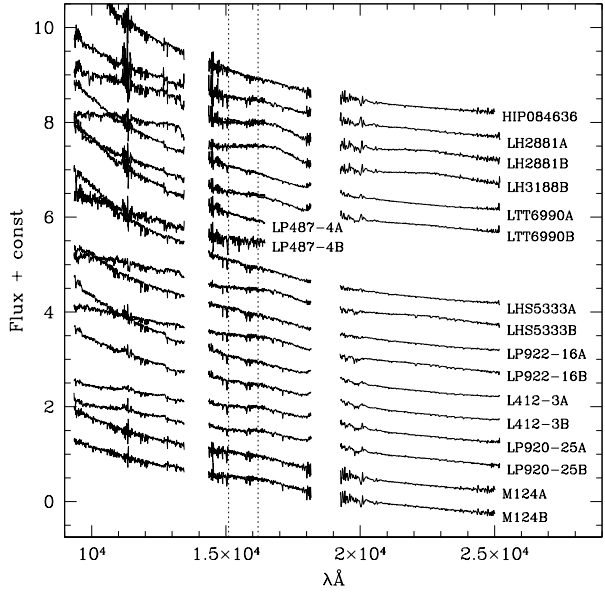}}
\caption{Near-infrared spectra of our targets. The spectral areas 
with poor atmospheric transmission are omitted. The spectra were 
normalized to 0.5 in the overlapping region (bracketed with dotted 
lines) and shifted vertically by 0.5 for clarity. M124A and M124B 
indicate MACHO\,124.22158.2900 and MACHO\,124.22158.2910, respectively, 
L\,412$-$3\,B is an alternative notation of HD\,322416 and LP\,920$-$25\,B
is the same for LP\,920$-$26. The spectrum of the telluric HIP\,084636 
shown here was corrected for the atmospheric absorption with 
HIP\,098813.}\label{fig:spectra_all}
\end{figure}

The data reduction steps were: sky/dark/bias removal by subtracting 
from each other the two complementary images in a nodding pair; 
flat fielding with dome flats; extraction of 1-dimensional (1-D) 
spectra from each star, on each individual image, by tracing the 
stellar continuum, with 6-8\,pixel (1\,pixel$\sim$0.29\,arcsec) 
wide apertures, with the IRAF\footnote{IRAF is distributed by the 
NOAO, which is operated by the AURA under cooperative agreement 
with NSF.} task {\it apall}; wavelength calibration of each 1-D 
stellar spectrum with 1-D Xenon lamp spectrum, extracted from Xenon 
lamp images with the same trace as each target spectrum; 
combination of the four or six 1-D spectra of each star in 
wavelength space with the IRAF task {\it scombine}, with 
appropriate masking or rejection of remaining detector artifacts 
and cosmic ray affected regions; telluric correction with spectra 
of near-solar analogs (G1V-G3V), observed just before or after 
the science target, at similar airmass, and reduced the same way; 
recovery of the original spectral shape and removal of the 
artificial emission lines \citep{1996AJ....111..537M} by 
multiplying with spectra of corresponding spectral type star from 
the flux-calibrated IRTF library \citep{2005ApJ...623.1115C,
2009ApJS..185..289R}. The signal-to-noise of the final spectra 
varies significantly with the target's brightness and with 
wavelength, but the areas clear from telluric absorption have 
S/N$\sim$10--30. The final spectra are plotted in 
Fig.\,\ref{fig:spectra_all}.

The spectral typing was performed comparing the overall shape of 
the SofI spectra with spectra from the IRTF library (Fig.\,3)
and the results are listed in 
Table\,\ref{table:spec_obs_log}. The typical uncertainty, 
estimated from a comparison with template stars of neighboring 
sub-types 
(Fig.\,3), is one sub-type. It 
was determined by comparing our targets with IRTF 
spectra of stars with nearby subtypes, and comparing stars with 
multiple IRTF observations. Finally, we corrected for telluric 
absorption the telluric standard HIP\,084636 with HIP\,098813, 
and re-determined its spectral type obtaining a best match with 
G2V star, to be compared with G3V reported by Gray et al. 
\citep{2006AJ....132..161G}.

\section{Discussion and Summary}\label{sec:results}

Why have the new HPM stars not been detected before? Some of them 
appear on old photographic surveys but the contamination from 
nearby stars, aggravated by the poor spatial resolution of those 
surveys, makes the identification of the stars as HPM objects 
difficult. The extreme differences between optical and infrared 
brightness of stars that is often found in the Galactic plane 
often led to misidentifications and some spurious HPM 
detections while true HPM stars were missed. Even with the 
high-quality of the VVV data we cannot consider the position of 
the stars reliable because of the uneven background. Multiple 
measurements are needed, separated by some years, to let the 
stars move by at least 2-3\,arcsec, so they lay on a completely 
different background, averaging out the contamination effects.

The PM errors in Table\,\ref{table:companions} are the r.m.s. for 
3-4 measurements (Table\,\ref{table:multi-epoch_coo}), and they 
only include statistical uncertainties. A comparison with the 
measurements in the literature suggest that the real uncertainties 
are larger. Excluding the obvious errors which yield differences 
exceeding 100\,mas\,yr$^{-1}$, $e.g.$ due to missidentification, 
we find an average difference of 2\,mas\,yr$^{-1}$, with an r.m.s 
of 17\,mas\,yr$^{-1}$, and suggest that the reader uses the latter 
number as the real error of our PMs that includes both internal 
and external uncertainties. The scatter gives an upper limit 
to the unaccounted bulk PM of the filed stars used for the 
astrometric calibration of the VVV data, and these are indeed 
small. More accurate measurements will become available in the 
future, as the VVV survey progresses. The planned survey duration 
of 5\,yrs is likely to be extended to $\sim$7\,yrs.

Some of the objects with overestimated PMs are very red. 
Interestingly, three of them were considered HPM objects despite 
being classified as Carbon stars, suggesting that they were giants. 
Probably, unaccounted astrometric color terms, combined with the 
extreme colors, have led to the erroneous classification. 

Notes on some individual objects:

-- LHS\,2881\,B is $\sim$8.1\,arcsec away from a HPM object listed 
in Monet et al. \citep{2003AJ....125..984M} with $\mu$(RA)=198$\pm$38 
and $\mu$(Dec)=848$\pm$319\,arcsec\,yr$^{-1}$ which is absent on our
data and it is likely a result of a missidentification or a spurious
entry in the USNOB1.0. Interestingly, the LHS\,2881 pair has 
similar PM to that of LHS\,2871: $\mu$(RA)=$-$461.01$\pm$1.67 and 
$\mu$(Dec)=$-$645.32$\pm$1.31\,arcsec\,yr$^{-1}$ as reported by van 
Leeuwen \citep{2007A&A...474..653V}. The wide separation of 
$\sim$44 arcmin makes it unlikely that they are bound but may 
indicate a common origin.

-- LP\,487-4 is projected on the sky close to the open cluster 
NGC\,6475 (M7) but it is not a physical member because the 
cluster has $\mu$(RA)=2.58 and $\mu$(Dec)=$-$4.54\,arcsec\,yr$^{-1}$
\citep[Loktin \& Beshenov][]{2003ARep...47....6L}. Furthermore, the 
optical spectroscopy of James et al. \citep{2000ASPC..198..277J} 
yields a radial velocity V$_{\rm rad}$=78.6$\pm$0.2\,km\,s$^{-1}$, 
inconsistent with V$_{\rm rad}$=$-$14.21$\pm$1.39\,km\,s$^{-1}$ of
NGC\,6475 \citep{2005A&A...438.1163K}.

-- LTT\,5140\,A parameters were derived from optical spectroscopy and 
Str\"omgren photometry by Nordstr\"om et al. \citep{2004A&A...418..989N}: 
$log$\,T$_{\rm eff}$=3.785, [Fe/H]=0.04, M$_V$=3.67\,mag, Age=3.3\,Gyr 
V$_{\rm rad}$=15.9$\pm$0.2\,km/s. Later, Holmberg et al. 
\citep{2009A&A...501..941H} updated them to $log$\,T$_{\rm eff}$=3.774, 
[Fe/H]=$-$0.06, and M$_V$=3.63\,mag to reflect the revised HIPPARCOS 
parallaxes. Desidera et al. \citep{2006A&A...454..553D} estimated from 
chromospheric activity $log$\,age=9.82 and 9.58 for the primary and 
the secondary, respectively.

-- LTT\,7318 and LTT\,7319 were considered a binary by Dommanget 
\citep{1983BICDS..24...83D} but later measurement by Salim \& Gould 
\citep{2003ApJ...582.1011S}, and van Leeuwen \cite{2007A&A...474..653V}
indicate that the two stars are not physically connected. Our data 
support this conclusion.

-- Some objects were included in our sample just because one source,
namely Monet et al. \citep{2003AJ....125..984M} reported HPM for them, 
despite the fact that other works have estimated low PM. For example,
C*\,1925, C*\,1930, C*\,1936, which are known carbon stars, i.e. 
distant giants, as reported by Alksnis et al. \citep{2001BaltA..10....1A}.

-- CD$-$60\,4613 was considered a HPM star by Turon et al. 
\citep{1992ESASP1136.....T} but it was probably misidentified with 
the nearby LTT\,5126 because of the large error in the NLTT 
coordinates of the latter star, identified by Salim \& Gould
\citep{2003ApJ...582.1011S}. Indeed, van Leeuwen 
\citep{2007A&A...474..653V} reported correct position and low PM for 
this star in the revised HIPPARCOS catalog under HIP\,65056.

-- the HPMs for OGLE\,BUL$-$SC20\,625107, OGLE BUL$-$ SC21\,298351, 
OGLE\,BUL$-$SC32\,388121 are subject to various sources of 
systematics: blending, contamination from variable sources, and 
seeing variations, aggravated by the crowded OGLE fields
\citep[see for details Sec.\,7 in ][]{2004MNRAS.348.1439S}.

HPM stars are nearby objects, and finding seven of them implies an 
incompleteness of $\sim$4\% (over 167 HPM stars) in the Solar 
neighborhood census. However, this is not a firm limit because: 
(i) it refers only to the bright stars considered here, (ii) the 
starting list of 167 stars is likely incomplete itself, and 
(iii) it is contaminated by non-moving stars, as we showed. 
Therefore, we refrain from making statements on the completeness
of the Solar neighborhood census; we only demonstrated that the HPM 
census is lacking stars, and that high angular resolution surveys 
help for addressing this issue in the most crowded regions of 
the Galaxy.

The new generation of near-infrared surveys of the Milky Way will 
produce enormous amounts of data, allowing the possibility of many 
discoveries. This work allows us to refine the strategy for future 
surveys and HPM star searches in the densest regions of the Southern 
Milky Way disk and the Bulge. We expect that many more HPM stars and
companions to them--including brown dwarfs and even planetary mass 
objects--will be discovered by these surveys when the baseline of 
observations reaches a few years, contributing to complete the 
census of faint nearby stars, and their multiplicity.

\begin{acknowledgements}
We acknowledge support by the FONDAP Center for Astrophysics 15010003;  
BASAL CATA Center for Astrophysics and Associated Technologies PFB-06; 
the Ministry for the Economy, Development, and Tourism's Programa 
Iniciativa Cient\'{i}fica Milenio through grant P07-021-F, awarded to 
The Milky Way Millennium Nucleus; FONDECYT grants No. 1090213 and 
1110326 from CONICYT, and the European Southern Observatory. 
JCB acknowledge support from a Ph.D. Fellowship from CONICYT.
MG is financed by the GEMINI-CONICYT Fund, allocated to the project 
32110014. RK acknowledges partial support from FONDECYT through grant 
1130140. ELM acknowledges support from grant AyA2011-30147-C03-03;
JB acknowledge support from FONDECYT No. 1120601; 
ANC acknowledge support from GEMINI-CONICYT No. 32110005 and from 
Comitee Mixto ESO-GOBIERNO DE CHILE.
JAG acknowledge support from Proyecto Fondecyt Postdoctoral 3130552, 
Fondecyt Regular 1110326, and Anillos ACT-86
We gratefully acknowledge use of data from the ESO VISTA telescope, 
and data products from the Cambridge Astronomical Survey Unit. We have 
also made extensive use of the SIMBAD Database at CDS Strasbourg, of 
the 2MASS, which is a joint project of the University of Massachusetts 
and IPAC/CALTECH, funded by NASA and NSF, and of the VizieR catalogue 
access tool, CDS, Strasbourg, France.
Last but not least, we thank the anonymous referee for the thoughtful 
and helpful comments that greatly improved the paper.
\end{acknowledgements}

\clearpage

\clearpage

\clearpage

\begin{figure}
\resizebox{\hsize}{!}{\includegraphics{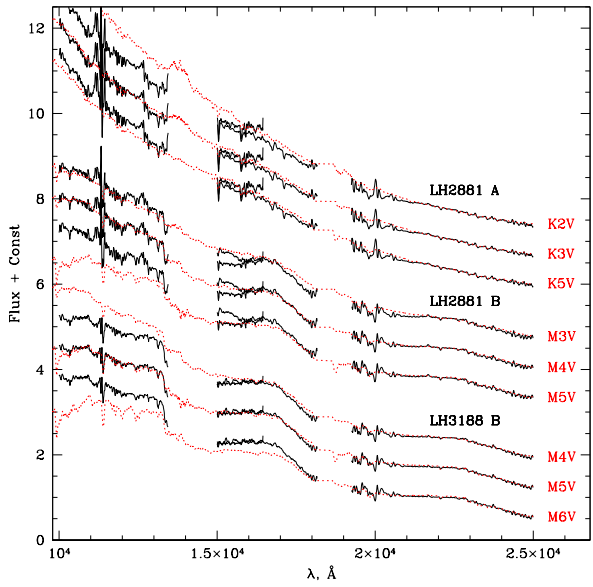}}
\caption{Spectral classification example. Solid lines show our 
spectra, and dotted lines--the template spectra from the IRTF 
library \citep{	2005ApJ...623.1115C,2009ApJS..185..289R}.}
\label{fig:spectra_class}
\end{figure}

\clearpage

\begin{table*}
\caption{HPM stars in the VVV area.}\label{table:targets} 
\begin{center}
\begin{tabular}{@{}l@{ }l@{ }l@{ }l@{ }l@{ }l@{ }l@{ }l@{}}
\hline\hline
Star ID & VVV\,Tiles~~ & Star ID & VVV\,Tiles~~ & Star ID & VVV\,Tiles~~ & Star ID & VVV\,Tiles \\
\hline
\multicolumn{8}{l}{} \\
\multicolumn{8}{c}{Disk area stars:} \\
\multicolumn{8}{l}{} \\
LTT\,18269                 & d007         & NLTT\,37871                & d053         & L\,200$-$41                & d095~~     & SCR\,J1440$-$5745~         & d130 \\      
LTT\,5140                  & d008         & HR\,5459                   & d053         & LSR\,J15292$-$5620         & d096       & L\,199$-$84                & d131 \\      
L\,149$-$77                & d012         & HR\,5460                   & d053         & L\,263$-$307               & d097       & SCR\,J1448$-$5735          & d131 \\      
V*\,V645\,Cen              & d014         & LSR\,J14570$-$5943         & d055         & LHS\,401                   & d098       & L\,263$-$307               & d135 \\      
LSR\,J14382$-$6231         & d014         & LSR\,J14585$-$5916         & d055         & L\,339$-$20                & d101       & L\,264$-$83                & d137 \\      
LTT\,6019                  & d017         & LTT\,5981                  & d055         & LTT\,6467                  & d102       & L\,264$-$78                & d138 \\      
L\,200$-$104               & d019         & L\,200$-$52                & d059         & LHS\,3185                  & d104       & LHS\,3182                  & d142 \\      
GJ\,9547                   & d025,\,d063  & SCR\,J1637$-$4703          & d068,\,d106~~& LHS\,3233                  & d107       & LHS\,3188                  & d143 \\      
LHS\,3223                  & d029         & LTT\,6763                  & d071         & LTT\,6714                  & d109       & L\,265$-$6                 & d140 \\      
LP\,413$-$57               & d036         & LTT\,6830                  & d073         & LTT\,6709                  & d109       & LTT\,6612                  & d144 \\      
C*\,1925                   & d039         & GJ\,662A                   & d075         & L\,412$-$30                & d111       & LTT\,6601                  & d144 \\      
C*\,1936                   & d040,\,d078~~& GJ\,662B                   & d075         & LP\,485$-$113              & d113       & L\,339$-$20                & d144 \\      
LHS\,2530                  & d041         & C*\,1930                   & d078         & LP\,485$-$65               & d114       & LHS\,3235                  & d146 \\      
TYC\,8994$-$252$-$1~       & d047         & LTT\,4656                  & d080         & LTT\,4656                  & d118       & L\,412$-$3                 & d149 \\      
NLTT\,35378                & d049         & SCR\,J1319$-$6200          & d084         & LTT\,5126                  & d122       & HD\,322416                 & d149 \\      
LHS\,2871                  & d050         & L\,148$-$26                & d085         & CD$-$60\,4613              & d122       & L\,412$-$30                & d149 \\      
LHS\,2881                  & d051         & L\,199$-$110               & d093         & LHS\,2892                  & d128       &                            &      \\
\multicolumn{8}{l}{} \\
\multicolumn{8}{c}{Bulge area stars:} \\
\multicolumn{8}{l}{} \\
LP\,487$-$82               & b203         & MACHO\,124.22028.40        & b264         & LTT\,7207                  & b295       & LHS\,443                   & b342        \\
V*\,eta\,Sgr               & b205         & MACHO\,124.22158.2900      & b264         & OGLE\,BUL$-$SC42\,133638~  & b295       & [TBF80]\,22                & b344        \\
NLTT\,46160                & b221         & MACHO\,124.22158.2910      & b264         & L\,486$-$29                & b300       & LP\,920$-$61               & b345        \\
LTT\,7388                  & b227         & MACHO\,124.22801.395       & b264         & LP\,486$-$36               & b301       & LP\,920$-$25               & b345        \\
LTT\,7182                  & b232         & MACHO\,116.24513.110       & b265         & LTT\,7072                  & b304       & LP\,920$-$26               & b345        \\
LP\,487$-$35               & b232,\,b246  & LP\,922$-$14               & b266         & LTT\,7073                  & b304       & NLTT\,45088                & b346        \\
LHS\,3372                  & b235         & LTT\,7259                  & b266         & LP\,866$-$12               & b309       & LP\,865$-$14               & b348        \\
NLTT\,46160                & b235         & OGLE\,BUL$-$SC29\,39202    & b273         & LP\,866$-$13               & b310       & LTT\,7150                  & b352        \\
LTT\,7233                  & b235         & LP\,559$-$192              & b274,\,b288  & LHS\,5333                  & b311       & G\,154$-$40                & b354        \\
LP\,922$-$20               & b237         & LP\,487$-$4                & b274         & NLTT\,45321                & b318       & NLTT\,44877                & b359        \\
LTT\,7318                  & b238         & LHS\,3335                  & b275         & LP\,921$-$21               & b319       & LP\,920$-$55               & b359        \\
LTT\,7319                  & b238         & LP\,559$-$120              & b276         & LP\,921$-$20               & b319       & LHS\,3310                  & b362        \\
LHS\,5337a                 & b239         & LP\,921$-$23               & b277         & LP\,921$-$15               & b320       & NLTT\,45440                & b365        \\
LP\,922$-$17               & b239         & LP\,921$-$26               & b277         & LP\,921$-$16               & b321       & LP\,864$-$17               & b376        \\
LHS\,3337                  & b244         & LTT\,6990                  & b285         & LTT\,7183                  & b324       & G\,154$-$29                & b381        \\
LP\,414$-$1                & b243         & LHS\,3330                  & b289         & LP\,558$-$60               & b330       & LP\,808$-$23               & b382,\,b396 \\
LP\,487$-$37               & b246         & LP\,921$-$28               & b291         & NLTT\,45128                & b331       & NLTT\,44500                & b386        \\
LP\,922$-$18               & b250         & OGLE\,BUL$-$SC20\,625107~  & b292         & NLTT\,45014                & b331       & [TBF80]\,10b               & b387        \\
LP\,866$-$17               & b253         & HD\,316899                 & b292         & NLTT\,45163                & b332       & [TBF80]\,10a               & b387        \\
LP\,922$-$15               & b253         & OGLE\,BUL$-$SC21\,298351   & b292         & LP\,921$-$11               & b333       & NLTT\,44488                & b387        \\
LP\,922$-$16               & b253         & OGLE\,BUL$-$SC32\,388121   & b293         & LP\,921$-$13               & b334       & LTT\,6900                  & b387        \\
LP\,866$-$19               & b254         & OGLE\,BUL$-$SC02\,783242   & b293         & LP\,865$-$20               & b336       & LTT\,6890                  & b387        \\
GJ\,2136 A                 & b255         & LTT\,7186                  & b294,\,b308  & LTT\,6873                  & b341       & LP\,864$-$16               & b389        \\
GJ\,2136 B                 & b255         & LP\,921$-$18               & b294         & LTT\,6869                  & b341       & LP\,864$-$14               & b390        \\
LP\,559$-$37               & b264         & NLTT\,45948                & b294         & HD\,156384                 & b342       & LTT\,7022                  & b392        \\
\hline
\end{tabular}
\end{center}
\end{table*}

\clearpage

\begin{table*}
\caption{Multi-epoch observations of target stars. Typical 
positional uncertainties:
$\sim$0.1\,arcsec for 2MASS, and
$\sim$0.05\,arcsec for VVV.}\label{table:multi-epoch_coo} 
\begin{center}
\begin{tabular}{@{}l@{ }c@{ }c@{ }c@{}}
\hline\hline
Image       & Primary                     & Secondary                   &~Epoch, UT           \\ 
Source      & RA Dec (J2000)              & RA Dec (J2000)              &~yyyy mm dd hh mm ss \\ 
\hline
\multicolumn{4}{c}{LTT\,5140 A -- LTT\,5140 B, separation $\sim$5\,arcsec} \\
2MASS       & 13 21 23.211 $-$64 02 59.08~&~13 21 23.965 $-$64 02 59.68 & 2000 04 27 02 57 44 \\ 
VVV\,$K_S$~ & 13 21 22.873 $-$64 02 59.51~&~13 21 23.657 $-$64 03 00.09 & 2010 02 19 08 02 07 \\ 
VVV\,$K_S$  & 13 21 22.882 $-$64 02 59.58~&~13 21 23.651 $-$64 03 00.20 & 2010 03 17 06 27 32 \\ 
VVV\,J      & 13 21 22.867 $-$64 02 59.56~&~13 21 23.636 $-$64 03 00.21 & 2010 03 17 06 36 38 \\ 
\multicolumn{4}{c}{L\,149$-$77 A -- L149$-$77 B, separation $\sim$14\,arcsec} \\
2MASS       & 14 12 28.089 $-$62 56 14.90~&~14 12 28.829 $-$62 56 27.86 & 2000 02 21 08 08 07 \\ 
VVV\,$K_S$  & 14 12 28.351 $-$62 56 12.84~&~14 12 29.088 $-$62 56 25.93 & 2010 03 05 07 28 16 \\ 
VVV\,$K_S$  & 14 12 28.313 $-$62 56 12.97~&~14 12 29.085 $-$62 56 25.94 & 2010 03 19 05 07 05 \\ 
VVV\,J      & 14 12 28.329 $-$62 56 12.91~&~14 12 29.086 $-$62 56 25.91 & 2010 03 19 05 15 03 \\ 
VVV\,$K_S$  & 14 12 28.323 $-$62 56 12.98~&~14 12 29.086 $-$62 56 25.91 & 2010 03 28 06 20 19 \\ 
\multicolumn{4}{c}{LHS\,2881 A -- LHS\,2881 B, separation $\sim$20\,arcsec} \\
2MASS       & 14 13 32.369 $-$62 07 33.37~&~14 13 30.335 $-$62 07 45.70 & 2000 02 21 08 15 58 \\ 
VVV\,J      & 14 13 31.584 $-$62 07 39.00~&~14 13 29.554 $-$62 07 51.21 & 2010 03 19 06 01 00 \\ 
VVV\,$K_S$  & 14 13 31.581 $-$62 07 38.95~&~14 13 29.555 $-$62 07 51.20 & 2010 03 19 05 54 07 \\ 
VVV\,$K_S$  & 14 13 31.479 $-$62 07 39.59~&~14 13 29.442 $-$62 07 51.94 & 2011 08 05 01 36 56 \\ 
\multicolumn{4}{c}{L\,200$-$41 A -- L\,200$-$41 B, separation $\sim$18\,arcsec} \\
2MASS       & 15 18 40.449 $-$56 27 55.70~&~15 18 39.953 $-$56 28 12.94 & 1999 06 07 05 21 16 \\ 
VVV\,J      & 15 18 40.240 $-$56 27 57.17~&~15 18 39.743 $-$56 28 14.43 & 2010 04 02 07 27 33 \\ 
\multicolumn{4}{c}{LHS\,3188 A -- LHS\,3188 B, separation $\sim$2.5\,arcsec} \\
2MASS       & 16 24 21.178 $-$46 44 01.78~&~16 24 21.202 $-$46 44 05.78 & 1999 05 20 03 43 14 \\ 
VVV\,J      & 16 24 20.651 $-$46 44 10.28~&~16 24 20.706 $-$46 44 12.87 & 2010 03 03 08 02 34 \\ 
VVV\,$K_S$  & 16 24 20.656 $-$46 44 09.95~&~16 24 20.705 $-$46 44 12.84 & 2010 03 03 07 55 46 \\ 
VVV\,$K_S$  & 16 24 20.598 $-$46 44 11.06~&~16 24 20.645 $-$46 44 13.75 & 2011 05 17 06 49 08 \\ 
\multicolumn{4}{c}{L412$-$3 -- L412$-$4 (HD\,322416), separation $\sim$26\,arcsec} \\
2MASS       & 16 58 45.218 $-$40 13 03.85~&~16 58 43.324 $-$40 13 18.64 & 1999 05 11 06 58 13 \\ 
VVV\,$K_S$  & 16 58 45.087 $-$40 13 05.16~&~16 58 43.193 $-$40 13 19.96 & 2010 03 26 08 48 03 \\ 
VVV\,J      & 16 58 45.091 $-$40 13 05.16~&~16 58 43.193 $-$40 13 19.97 & 2010 03 26 08 48 03 \\ 
\multicolumn{4}{c}{LP\,920-25 -- LP\,920-26, separation $\sim$8.5\,arcsec} \\
2MASS       & 17 31 40.047 $-$30 40 56.51~&~17 31 40.510 $-$30 41 02.62 & 1998 08 10 03 19 18 \\ 
VVV\,$K_S$  & 17 31 39.894 $-$30 40 58.87~&~17 31 40.360 $-$30 41 04.89 & 2010 04 15 08 56 31 \\ 
VVV\,$K_S$  & 17 31 39.884 $-$30 40 59.05~&~17 31 40.348 $-$30 41 05.19 & 2010 08 03 04 18 06 \\ 
VVV\,J      & 17 31 39.887 $-$30 40 58.97~&~17 31 40.353 $-$30 41 05.08 & 2010 08 03 04 23 02 \\ 
\multicolumn{4}{c}{LTT\,6990 A -- LTT\,6990 B, separation $\sim$3.5\,arcsec} \\
2MASS       & 17 35 08.353 $-$38 37 27.94~&~17 35 08.644 $-$38 37 29.89 & 2000 07 05 00 38 29 \\ 
VVV\,$K_S$  & 17 35 08.087 $-$38 37 30.51~&~17 35 08.374 $-$38 37 32.30 & 2010 08 31 02 58 30 \\ 
VVV\,J      & 17 35 08.080 $-$38 37 30.49~&~17 35 08.372 $-$38 37 32.15 & 2010 08 31 03 03 35 \\ 
VVV\,$K_S$  & 17 35 08.072 $-$38 37 30.70~&~17 35 08.351 $-$38 37 32.48 & 2011 07 27 04 43 36 \\ 
\multicolumn{4}{c}{LP\,487-4 A -- LP\,487-4 B, separation $\sim$36\,arcsec} \\
2MASS       & 17 51 22.238 $-$35 05 58.39~&~17 51 23.322 $-$35 05 25.88 & 1998 08 14 23 35 01 \\ 
VVV\,$K_S$  & 17 51 22.096 $-$35 05 59.79~&~17 51 23.180 $-$35 05 27.25 & 2010 04 21 09 24 05 \\ 
VVV\,J      & 17 51 22.091 $-$35 05 59.81~&~17 51 23.175 $-$35 05 27.28 & 2010 04 21 09 27 24 \\ 
VVV\,$K_S$  & 17 51 22.090 $-$35 05 59.84~&~17 51 23.173 $-$35 05 27.29 & 2010 06 27 07 58 12 \\ 
\multicolumn{4}{c}{MACHO\,124.22158.2900 -- MACHO\,124.22158.2910, separation $\sim$3\,arcsec} \\
2MASS       & 18 07 57.462 $-$30 54 55.50~&~18 07 57.313 $-$30 54 58.14 & 1998 07 27 01 31 25 \\ 
VVV\,$K_S$  & 18 07 57.323 $-$30 54 58.46~&~18 07 57.172 $-$30 55 01.13 & 2010 08 15 04 04 01 \\ 
VVV\,J      & 18 07 57.324 $-$30 54 58.43~&~18 07 57.171 $-$30 55 01.07 & 2010 08 15 04 05 45 \\ 
\multicolumn{4}{c}{LHS\,5333 A -- LHS\,5333 B, separation $\sim$43\,arcsec} \\
2MASS       & 18 09 17.677 $-$22 54 30.03~&~18 09 18.212 $-$22 55 12.06 & 1999 07 07 01 49 09 \\ 
VVV\,$K_S$  & 18 09 17.713 $-$22 54 35.12~&~18 09 18.243 $-$22 55 16.95 & 2010 04 21 06 10 06 \\ 
VVV\,J      & 18 09 17.708 $-$22 54 35.23~&~18 09 18.245 $-$22 55 16.95 & 2010 04 21 06 13 16 \\ 
VVV\,$K_S$  & 18 09 17.713 $-$22 54 35.24~&~18 09 18.240 $-$22 55 17.13 & 2010 10 06 01 38 42 \\ 
\multicolumn{4}{c}{LP\,922$-$16 A -- LP922$-$16 B, separation $\sim$95\,arcsec} \\
2MASS       & 18 23 51.140 $-$27 46 18.21~&~18 23 52.896 $-$27 47 50.87 & 1998 07 19 02 56 37 \\ 
VVV\,J      & 18 23 51.029 $-$27 46 20.65~&~18 23 52.785 $-$27 47 53.29 & 2010 04 08 07 35 20 \\ 
VVV\,$K_S$  & 18 23 51.029 $-$27 46 20.72~&~18 23 52.784 $-$27 47 53.26 & 2010 04 08 07 32 05 \\ 
VVV\,$K_S$  & 18 23 51.026 $-$27 46 20.81~&~18 23 52.776 $-$27 47 53.38 & 2010 10 26 23 57 33 \\ 
\multicolumn{4}{c}{LTT\,7318 -- LTT\,7319, separation $\sim$46\,arcsec} \\
2MASS       & 18 24 26.943 $-$29 32 39.75~&~18 24 26.552 $-$29 31 54.45 & 1998 07 19 03 10 49 \\ 
VVV\,$K_S$  & 18 24 27.119 $-$29 32 41.40~&~18 24 26.560 $-$29 31 56.99 & 2010 08 15 02 34 29 \\ 
VVV\,J      & 18 24 27.124 $-$29 32 41.24~&~18 24 26.556 $-$29 31 56.94 & 2010 08 15 02 38 06 \\ 
VVV\,$K_S$  & 18 24 27.135 $-$29 32 41.40~&~18 24 26.563 $-$29 31 56.94 & 2010 08 15 02 38 06 \\ 
\multicolumn{4}{c}{GJ\,2136 A -- GJ\,2136 B, separation $\sim$21\,arcsec} \\
2MASS       & 18 27 18.623 $-$25 04 23.59~&~18 27 18.432 $-$25 04 02.55 & 1998 07 19 03 46 39 \\ 
VVV\,$K_S$  & 18 27 18.543 $-$25 04 25.78~&~18 27 18.362 $-$25 04 04.78 & 2010 04 08 08 41 39 \\ 
VVV\,J      & 18 27 18.543 $-$25 04 25.71~&~18 27 18.362 $-$25 04 04.73 & 2010 09 05 12 17 24 \\ 
VVV\,$K_S$  & 18 27 18.535 $-$25 04 25.76~&~18 27 18.355 $-$25 04 04.80 & 2010 10 26 00 33 08 \\ 
\hline
\end{tabular}
\end{center}
\end{table*}

\clearpage

\begin{table}
\caption{Multi-epoch observations of stars with over-estimated PMs. 
Typical positional uncertainties:
$\sim$0.1\,arcsec for 2MASS, and
$\sim$0.05\,arcsec for VVV.}\label{table:multi-epoch_coo-not-moving} 
\begin{center}
\begin{tabular}{@{}l@{ }c@{ }c@{}}
\hline\hline
Image       & Coordinates                 & Epoch, UT                \\ 
Source      & RA Dec (J2000)              & yyyy\,mm\,dd\,hh\,mm\,ss \\ 
\hline
\multicolumn{3}{c}{C*\,1925} \\
2MASS       & 11 52 11.709 $-$62 15 00.18 & 2000 02 14 04 28 32 \\ 
VVV\,$K_S$  & 11 52 11.700 $-$62 14 59.79 & 2010 03 15 03 35 33 \\ 
VVV\,J      & 11 52 11.686 $-$62 14 59.91 & 2010 03 15 03 42 15 \\ 
VVV\,$K_S$  & 11 52 11.721 $-$62 14 59.99 & 2011 07 24 23 25 52 \\ 
\multicolumn{3}{c}{C*\,1930} \\
2MASS       & 11 54 49.494 $-$61 30 54.03 & 2000 02 14 04 43 24 \\ 
VVV\,$K_S$  & 11 54 49.495 $-$61 30 54.13 & 2010 03 14 03 29 49 \\ 
VVV\,J      & 11 54 49.509 $-$61 30 54.09 & 2010 03 14 03 36 38 \\ 
VVV\,$K_S$  & 11 54 49.518 $-$61 30 54.18 & 2011 06 12 02 44 58 \\ 
\multicolumn{3}{c}{C*\,1936} \\
2MASS       & 11 56 55.808 $-$62 15 30.93 & 2000 02 14 05 13 14 \\ 
VVV\,$K_S$  & 11 56 55.798 $-$62 15 31.11 & 2010 03 15 03 56 04 \\ 
VVV\,J      & 11 56 55.818 $-$62 15 30.87 & 2010 03 15 04 03 43 \\ 
VVV\,$K_S$  & 11 56 55.808 $-$62 15 31.01 & 2011 05 20 23 45 27 \\ 
\multicolumn{3}{c}{CD$-$60\,4613} \\
2MASS       & 13 20 07.358 $-$61 29 34.65 & 2000 04 17 02 41 08 \\ 
VVV\,$K_S$  & 13 20 07.367 $-$61 29 34.73 & 2010 03 07 09 01 48 \\ 
VVV\,J      & 13 20 07.347 $-$61 29 34.66 & 2010 03 07 09 08 40 \\ 
VVV\,$K_S$  & 13 20 07.331 $-$61 29 34.81 & 2011 06 14 03 38 57 \\ 
\multicolumn{3}{c}{LP\,866$-$17} \\
2MASS       & 18 20 55.854 $-$27 05 55.09 & 1998 07 19 02 14 03 \\ 
VVV\,$K_S$  & 18 20 55.851 $-$27 05 55.13 & 2010 04 08 07 32 05 \\ 
VVV\,J      & 18 20 55.851 $-$27 05 55.09 & 2010 04 08 07 35 20 \\ 
VVV\,$K_S$  & 18 20 55.849 $-$27 05 55.17 & 2010 10 26 23 57 33 \\ 
\multicolumn{3}{c}{OGLE\,BUL$-$SC20 625107} \\
2MASS       & 17 59 35.685 $-$29 11 57.34 & 1998 07 16 05 34 29 \\ 
VVV\,$K_S$  & 17 59 35.526 $-$29 11 58.82 & 2011 05 09 05 22 40 \\ 
VVV\,J      & 17 59 35.523 $-$29 11 58.84 & 2011 05 09 05 26 50 \\ 
VVV\,$K_S$  & 17 59 35.521 $-$29 11 58.89 & 2011 05 18 07 44 05 \\ 
\multicolumn{3}{c}{OGLE\,BUL$-$SC21 298351} \\
2MASS       & 18 00 12.122 $-$29 03 46.87 & 1999 07 05 03 48 58 \\ 
VVV\,$K_S$  & 18 00 12.120 $-$29 03 46.98 & 2011 05 09 05 22 40 \\ 
VVV\,J      & 18 00 12.118 $-$29 03 46.97 & 2011 05 09 05 26 50 \\ 
VVV\,$K_S$  & 18 00 12.116 $-$29 03 47.01 & 2011 05 18 07 44 05 \\ 
\multicolumn{3}{c}{OGLE\,BUL$-$SC32 388121} \\  
2MASS       & 18 03 11.721 $-$28 13 22.08 & 1998 03 19 09 21 19 \\ 
VVV\,$K_S$  & 18 03 11.733 $-$28 13 22.01 & 2011 05 09 05 51 15 \\ 
VVV\,J      & 18 03 11.729 $-$28 13 22.04 & 2011 05 09 05 55 12 \\ 
\hline
\end{tabular}
\end{center}
\end{table}

\clearpage

\begin{table*}
\caption[]{Measured PMs for new, known, and rejected co-moving pairs of stars.
The letter M in the star ID column stands for MACHO. PMs, parallaxes and spectral
types from the literature are also listed.\label{table:companions}}
\centering
\begin{tabular}{@{}l@{}r@{}r@{}r@{}r@{}c@{}c@{}c@{}c@{}c@{}c@{}c@{}}
\hline\hline
Star ID & \multicolumn{1}{c}{~~RA} & \multicolumn{1}{c}{~~Dec} & \multicolumn{2}{c}{~~PM\,(RA,Dec)} & \multicolumn{2}{c}{~~PM\,(RA,Dec)}    & Ref. & Parallax~&~Ref.~&~Sp.\,Type  & Ref. \\
        & \multicolumn{2}{c}{(2MASS)} & \multicolumn{2}{c}{~~[mas\,yr$^{-1}$], this work} & \multicolumn{2}{c}{~~[mas\,yr$^{-1}$], literature} & & [mas]      &      &~literature &  \\
\hline
\multicolumn{6}{l}{} \\
\multicolumn{6}{l}{New co-moving companions around known HPM stars:} \\
\multicolumn{6}{l}{} \\
L\,149-77 A          & 14 12 28.089 &~$-$62 56 14.90  &    162.5$\pm$11.9~&    196.3$\pm$6.7  &    154.8$\pm$4.5   &    184.3$\pm$4.1   &  (4) &~25.6$^{+12.8}_{-10.3}$~& (17) & K7V     & (17)/(18) \\ 
                     &              &                 &                   &                   &    158.1$\pm$3.1   &    183.5$\pm$2.9   & (15) & 20             & (19) & K5V       & (19) \\ 
                     &              &                 &                   &                   &    157.2$\pm$3.4   &    187.9$\pm$3.1   & (16) &~14.17$\pm$1.56~& (20) &           &     \\ 
                     &              &                 &                   &                   &    141.3$\pm$10.0  &    163.0$\pm$10.0  & (20) &                &      &           &     \\ 
L\,149-77 B          & 14 12 28.829 & $-$62 56 27.86  &    171.2$\pm$0.4  &    192.6$\pm$1.2  &    242.0$\pm$6.0   &    216.0$\pm$49.0  & (15) &                &      &           &     \\ 
                     &              &                 &                   &                   &    193.5$\pm$7.4   &    143.7$\pm$7.4   & (16) &                &      &           &     \\ 
\multicolumn{6}{l}{} \\
LHS\,2881 A          & 14 13 32.369 & $-$62 07 33.37  &~$-$541.1$\pm$1.5  & $-$551.7$\pm$7.9  &   $-$440$\pm$100   &   $-$607$\pm$100   &  (9) &                &      &           &     \\
                     &              &                 &                   &                   & $-$550.1$\pm$12.5  & $-$548.7$\pm$12.5  & (16) &                &      &           &     \\ 
                     &              &                 &                   &                   &   $-$496$\pm$19    &   $-$588$\pm$33    & (21) &                &      &           &     \\ 
                     &              &                 &                   &                   &    548.3$\pm$8.0   & $-$511.7$\pm$8.0   & (22) &                &      &           &     \\ 
LHS\,2881 B          & 14 13 30.335 & $-$62 07 45.70  & $-$545.8$\pm$1.0  & $-$545.3$\pm$1.1  &                    &                    &      &                &      &           &     \\ 
\multicolumn{6}{l}{} \\
L\,200-41 A          & 15 18 40.449 & $-$56 27 55.70  & $-$162.7$\pm$---- & $-$134.5$\pm$---- & $-$174.2$\pm$3.4   & $-$142.8$\pm$3.2   &  (4) & 24.47$\pm$11.91 & (20) &           &     \\ 
                     &              &                 &                   &                   & $-$185.2$\pm$3.5   & $-$143.9$\pm$1.3   & (15) &                &      &           &     \\ 
                     &              &                 &                   &                   & $-$175.00$\pm$2.7  & $-$141.47$\pm$2.7  & (16) &                &      &           &     \\ 
                     &              &                 &                   &                   & $-$132.8$\pm$10.0  & $-$108.5$\pm$10.0  & (20) &                &      &           &     \\ 
                     &              &                 &                   &                   & $-$176             & $-$144             & (21) &                &      &           &     \\ 
                     &              &                 &                   &                   & $-$172.8$\pm$1.2   & $-$139.7$\pm$1.2   & (22) &                &      &           &     \\ 
L\,200-41 B          & 15 18 39.953 & $-$56 28 12.94  & $-$159.8$\pm$---- & $-$138.3$\pm$---- & $-$180$\pm$22      & $-$126$\pm$36      & (21) &                &      &           &     \\ 
\multicolumn{6}{l}{} \\
LHS\,3188 A          & 16 24 21.178 & $-$46 44 01.78  & $-$491.3$\pm$1.4  &~$-$774.1$\pm$15.3 & $-$511.6           & $-$741.6           &~(12)~&  48$\pm$12    &~(10)~& K5       & (11) \\
                     &              &                 &                   &                   & $-$458$\pm$76      & $-$688$\pm$123     & (21) &                &      &           &     \\ 
                     &              &                 &                   &                   & $-$462.3$\pm$12.2  & $-$750.6$\pm$12.2  & (23) &                &      &           &     \\ 
                     &              &                 &                   &                   & $-$520             & $-$700             & (24) &                &      &           &     \\ 
LHS\,3188 B          & 16 24 21.202 & $-$46 44 05.78  & $-$478.9$\pm$3.7  & $-$658.5$\pm$5.1  &                    &                    &      &                &      &           &     \\ 
\multicolumn{6}{l}{} \\
LP\,487-4 A          & 17 51 22.238 & $-$35 05 58.39  & $-$165.8$\pm$3.9  &~$-$120.8$\pm$0.8  & $-$157.9$\pm$4.4   & $-$146.0$\pm$4.4   &  (1) &                &      &           &     \\ 
                     &              &                 &                   &                   & $-$158.35$\pm$2.7  & $-$136.33$\pm$2.7  & (16) &                &      &           &     \\ 
LP\,487-4 B          & 17 51 23.322 & $-$35 05 25.88  & $-$150.5$\pm$3.9  & $-$117.7$\pm$1.2  &                    &                    &      &                &      &           &     \\ 
\multicolumn{6}{l}{} \\
LHS\,5333 A          & 18 09 17.677 & $-$22 54 30.03  &     61.6$\pm$4.8  & $-$473.3$\pm$9.5  &   40.08$\pm$0.91   & $-$459.58$\pm$0.59 &  (2) &~31.54$\pm$0.93~&  (2) & K1IV      & (13) \\
                     &              &                 &                   &                   &   42.18$\pm$1.08   & $-$459.56$\pm$0.70 & (30) & 31.27$\pm$1.12 & (30) &           &     \\ 
                     &              &                 &                   &                   &   42.1$\pm$6.4     & $-$459.0$\pm$2.2   &  (1) &                &      &           &     \\ 
                     &              &                 &                   &                   &   46.8             & $-$477.7           & (12) &                &      &           &     \\ 
                     &              &                 &                   &                   &   42.1$\pm$1.0     & $-$459.5$\pm$0.6   & (15) &                &      &           &     \\ 
                     &              &                 &                   &                   &   42.1$\pm$1.0     & $-$459.5$\pm$0.7   & (22) &                &      &           &     \\ 
                     &              &                 &                   &                   &   37.7$\pm$6.4     & $-$458.7$\pm$8.2   & (23) &                &      &           &     \\ 
                     &              &                 &                   &                   &   55.6             & $-$489.7           & (25) &                &      &           &     \\ 
                     &              &                 &                   &                   &   45.9$\pm$3.1     & $-$461.5$\pm$3.1   & (26) &                &      &           &     \\ 
                     &              &                 &                   &                   &   43$\pm$18        & $-$486$\pm$18      & (27) &                &      &           &     \\ 
                     &              &                 &                   &                   &   52               & $-$409             & (28) &                &      &           &     \\ 
                     &              &                 &                   &                   &   42$\pm$2.9       & $-$462$\pm$2.9     & (29) &                &      &           &     \\ 
                     &              &                 &                   &                   &   44.64$\pm$2.48   & $-$458.9$\pm$3.92  & (31) &                &      &           &     \\ 

LHS\,5333 B          & 18 09 18.212 & $-$22 55 12.06  &     52.5$\pm$6.5  & $-$452.9$\pm$1.5  &                    &                    &      &                &      &           &     \\ 
\multicolumn{6}{l}{} \\
LP\,922-16 A         & 18 23 51.140 & $-$27 46 18.21  & $-$117.0$\pm$1.6  & $-$212.1$\pm$3.3  & $-$124.87$\pm$3.06 & $-$205.72$\pm$1.89 &  (2) &~10.93$\pm$2.29~& (2)  &           &     \\ 
                     &              &                 &                   &                   & $-$127.64$\pm$2.52 & $-$207.02$\pm$1.68 & (30) & 12.30$\pm$2.22 & (30) &           &     \\ 
                     &              &                 &                   &                   & $-$129.8$\pm$2.7   & $-$217.2$\pm$2.5   &  (1) &                &      &           &     \\ 
                     &              &                 &                   &                   & $-$127.6$\pm$2.5   & $-$207.0$\pm$1.6   & (15) &                &      &           &     \\ 
                     &              &                 &                   &                   & $-$129.39$\pm$1.9  & $-$210.12$\pm$1.4  & (16) &                &      &           &     \\ 
                     &              &                 &                   &                   & $-$130             & $-$218             & (21) &                &      &           &     \\ 
                     &              &                 &                   &                   & $-$128.6$\pm$2.4   & $-$209,8$\pm$2.2   & (22) &                &      &           &     \\ 
                     &              &                 &                   &                   & $-$126.1$\pm$6.5   & $-$217.0$\pm$6.0   & (32) &                &      &           &     \\ 
LP\,922-16 B         & 18 23 52.896 & $-$27 47 50.87  & $-$130.7$\pm$3.4  & $-$205.4$\pm$1.3  & $-$122.3$\pm$9.0   & $-$197.4$\pm$9.0   & (15) &                &      &           &     \\ 
                     &              &                 &                   &                   & $-$160$\pm$48      & $-$134$\pm$57      & (21) &                &      &           &     \\ 
                     &              &                 &                   &                   & $-$154.8$\pm$9.5   & $-$143.4$\pm$9.5   & (23) &                &      &           &     \\ 
\hline
\end{tabular}
\end{table*}

\addtocounter{table}{-1}
\begin{table*}
\caption[]{Continued.} 
\centering
\begin{tabular}{@{}l@{}r@{}r@{}r@{}r@{}c@{}c@{}c@{}c@{}c@{}c@{}c@{}}
\hline\hline
Star ID & \multicolumn{1}{c}{~~RA} & \multicolumn{1}{c}{~~Dec} & \multicolumn{2}{c}{~~PM\,(RA,Dec)} & \multicolumn{2}{c}{~~PM\,(RA,Dec)}    & Ref. & Parallax~&~Ref.~&~Sp.\,Type  & Ref. \\
        & \multicolumn{2}{c}{(2MASS)} & \multicolumn{2}{c}{~~[mas\,yr$^{-1}$], this work} & \multicolumn{2}{c}{~~[mas\,yr$^{-1}$], literature} & & [mas]      &      &~literature &  \\
\hline
\multicolumn{6}{l}{} \\
\multicolumn{6}{l}{Recovered known binaries:} \\                                  
\multicolumn{6}{l}{} \\                                                
LTT\,5140 A          & 13 21 23.965 & $-$64 02 59.68  & $-$223.4$\pm$6.3  &  $-$48.3$\pm$3.0  & $-$233.7$\pm$0.6   &   $-$44.0$\pm$0.8  &  (2) &~17.81$\pm$0.98~& (2)  & G0V       & (6) \\ 
                     &              &                 &                   &                   & $-$236.5$\pm$2.3   &   $-$47.6$\pm$2.3  &  (4) &                &      &           &     \\
LTT\,5140 B          & 13 21 23.211 & $-$64 02 59.08  & $-$207.1$\pm$8.1  &  $-$49.2$\pm$6.2  & $-$223.0$\pm$3.1   &   $-$47.0$\pm$3.0  &  (4) &                &      & G0        & (6) \\ 	
\multicolumn{6}{l}{} \\
L\,412-3             & 16 58 45.218 & $-$40 13 03.85  & $-$135.1$\pm$---- & $-$121.2$\pm$---- & $-$147.4$\pm$5.4   &  $-$136.0$\pm$5.4  &  (1) &                &      & K2        & (3) \\ 
L\,412-4             & 16 58 43.324 & $-$40 13 18.64  & $-$150.4$\pm$---- & $-$121.8$\pm$---- & $-$145.1$\pm$4.5   &  $-$129.0$\pm$4.5  &  (1) &                &      & K2        & (3) \\ 
\multicolumn{1}{r}{(= HD\,322416)} & &                &                   &                   & $-$144.3$\pm$2.2   &  $-$121.4$\pm$2.3  &  (4) &                &      &           &     \\
\multicolumn{6}{l}{} \\
LP\,920-25           & 17 31 40.047 &~$-$30 40 56.51  &~$-$164.3$\pm$2.2  & $-$206.5$\pm$5.0  & $-$181.7$\pm$2.0   &  $-$197.0$\pm$2.0  &  (1) &                &      &           &     \\ 
LP\,920-26           & 17 31 40.510 & $-$30 41 02.62  & $-$169.7$\pm$3.9  &~$-$215.6$\pm$10.2~& $-$181.7$\pm$2.0   &  $-$197.0$\pm$2.0  &  (1) &                &      &           &     \\ 
\multicolumn{6}{l}{} \\
LTT\,6990 A          & 17 35 08.353 & $-$38 37 28.15  & $-$295.7$\pm$8.6  & $-$250.9$\pm$1.8  &~$-$311.72$\pm$1.71~&~$-$228.28$\pm$1.08~&  (2) &~30.40$\pm$1.36~& (2)  & K0V       & (13) \\
LTT\,6990 B          & 17 35 08.644 & $-$38 37 29.89  & $-$317.4$\pm$4.3  & $-$230.6$\pm$7.8  &                    &                    &      &                &      &           &     \\ 
\multicolumn{6}{l}{} \\                                                      
M\,124.22158.2900~   & 18 07 57.462 & $-$30 54 55.50  & $-$148.6$\pm$---- & $-$244.7$\pm$---- & $-$119.7$\pm$5.0   &  $-$249.8$\pm$5.0  &  (5) &                &      &           &     \\ 
M\,124.22158.2910    & 18 07 57.313 & $-$30 54 58.14  & $-$140.8$\pm$---- & $-$245.9$\pm$---- & $-$115.8$\pm$5.0   &  $-$242.8$\pm$5.0  &  (5) &                &      &           &     \\ 
\multicolumn{6}{l}{} \\
GJ\,2136 A           & 18 27 18.623 & $-$25 04 23.59  &  $-$79.8$\pm$8.2  & $-$179.2$\pm$6.3  &  $-$84.4$\pm$3.2   &  $-$198.1$\pm$3.0  &  (1) &~30.86$\pm$4.57~& (7)  & M0/M0.5   & (14)/(7) \\ 
GJ\,2136 B           & 18 27 18.432 & $-$25 04 02.55  &  $-$77.1$\pm$3.7  & $-$184.7$\pm$5.3  &  $-$88.0$\pm$25.0  &  $-$168.0$\pm$25.0 &  (8) &                &      &           &     \\ 
\multicolumn{6}{l}{} \\
\multicolumn{6}{l}{Not co-moving:} \\
\multicolumn{6}{l}{} \\
LTT\,7318            & 18 24 26.943 & $-$29 32 39.75  &   184.7$\pm$12.1  & $-$132.1$\pm$7.6  &    194.4$\pm$2.0   &  $-$120.8$\pm$1.8  &  (1) &~21.30$\pm$1.08~& (2) & F8V       & (6) \\ 
                     &              &                 &                   &                   &    192.4$\pm$1.3   &  $-$120.3$\pm$0.8  &  (2) &                &     &           &     \\
LTT\,7319            & 18 24 26.559 & $-$29 31 54.53  &     0.0$\pm$7.9   & $-$207.4$\pm$2.6  &      0.5$\pm$1.5   &  $-$197.8$\pm$2.0  &  (1) &~21.38$\pm$0.92~& (2)~&~K3/K4III+ & (6) \\ 
                     &              &                 &                   &                   &   $-$1.1$\pm$1.1   &  $-$198.3$\pm$0.7  &  (2) &                &     &           &     \\
\hline
\end{tabular}
\tablebib{
(1) Salim \& Gould \citep{2003ApJ...582.1011S};
(2) van Leeuwen \citep{2007A&A...474..653V};
(3) Nesterov et al. \citep{1995A&AS..110..367N};
(4) H\/og et al. \citep{2000A&A...355L..27H};
(5) Alcock et al. \citep{2001ApJ...562..337A};
(6) as given in SIMBAD, the original reference is unknown;
(7) Reid et al. \citep{2004AJ....128..463R};
(8) Luyten \citep{1963BPM...C......0L};
(9) Bakos et al. \citep{2002ApJS..141..187B};
(10) Jenkins \citep{1963gcts.book.....J};
(11) Bidelman \citep{1985ApJS...59..197B};
(12) Luyten \citep{1979lccs.book.....L};
(13) Gray et al. \citep{2006AJ....132..161G};
(14) Stephenson \& Sanduleak \citep{1975AJ.....80..972S};
(15) Zacharias et al. \citep{2004AJ....127.3043Z};
(16) R\"oser et al. \citep{2008A&A...488..401R};
(17) Ammons et al. \citep{2006ApJ...638.1004A};
(18) Schmidt-Kaler \citep{SchK82};
(19) Pickles \& Depagne \citep{2010PASP..122.1437P};
(20) Fresneau et al. \citep{2007A&A...469.1221F};
(21) Monet et al. \citep{2003AJ....125..984M};
(22) Zacharias et al. \citep{2012yCat.1322....0Z} -- possibly, a wrong sign for the RA PM of LHS\,2881\,A;
(23) R\"oser et al. \citep{2010AJ....139.2440R};
(24) Stauffer et al. \citep{2010PASP..122..885S};
(25) Hog \& von der Heide \citep{1976AAHam...9....1H};
(26) Dunham \citep{Dunham1986};
(27) Smithsonian Astrophysical Observatory Star Catalog;
(28) Schlesinger \& Barney \citep{1943TOYal..14....1S};
(29) Bastian \& R\"oser \citep{Bastian1993};
(30) Perryman et al. \citep{1997ESASP1200.....P};
(31) Urban et al. \citep{Urban1997};
(32) R\"oser et al. \citep{1994A&AS..105..301R};
}
\end{table*}

\begin{table*}
\caption{List of stars with overestimated PMs. The first 
literature PM is the one listed in SIMBAD at the time of 
the sample preparation.}\label{table:not_high_pm} 
\begin{center}
\begin{tabular}{@{}l@{ }c@{ }c@{ }c@{ }c@{}c@{}c@{}c@{}}
\hline\hline
Star ID & \multicolumn{2}{c}{PM(RA,Dec)}                  & \multicolumn{2}{c}{PM(RA,Dec)}                   &~Ref.~& Sp.  &~Ref. \\ 
        & \multicolumn{2}{c}{[mas\,yr$^{-1}$], this work} & \multicolumn{2}{c}{[mas\,yr$^{-1}$], literature} &      & Type &      \\ 
\hline
C*\,1925                 & $-$1.3$\pm$12.0  & 25.6$\pm$10.6    & 256$\pm$110       & 342$\pm$40          & (1)  &~Carbon star~& (2) \\ 
C*\,1930                 & 10.6$\pm$7.1     & $-$10.3$\pm$3.3  &  18$\pm$91        & 278$\pm$46          & (1)  &~Carbon star~& (2) \\ 
C*\,1936                 & 5.9$\pm$7.6      & $-$6.5$\pm$11.6  & 146$\pm$44        & 226$\pm$7           & (1)  &~Carbon star~& (2) \\ 
CD$-$60\,4613            & $-$3.7$\pm$13.6  & $-$8.2$\pm$6.7   & $-$251$\pm$25     & $-$23$\pm$25        & (3)  &             &     \\ 
                         &                  &                  & $-$8.65$\pm$6.92  & $-$8.27$\pm$6.40    & (4)  &             &     \\ 
                         &                  &                  & $-$7.23$\pm$2.2   & $-$0.35$\pm$2.2     & (5)  &             &     \\ 
                         &                  &                  & $-$7.38$\pm$1.07  & $-$1.76$\pm$1.04    & (9)  &             &     \\
LP\,866$-$17             & $-$8.2$\pm$0.2   & $-$3.4$\pm$3.4   & $-$7.4$\pm$5.5    & $-$216.9$\pm$5.5    & (6)  &             &     \\ 
                         &                  &                  & $-$16.1$\pm$4.4   & $-$19.1$\pm$4.3     & (5)  &             &     \\ 
OGLE\,BUL$-$SC20 625107  &~$-$162.0$\pm$4.2 & $-$117.5$\pm$2.6~&~$-$42.56$\pm$6.96~&~$-$330.93$\pm$32.81~& (7)  &             &     \\ 
OGLE\,BUL$-$SC21 298351~~& 8.1$\pm$1.5      & $-$10.4$\pm$1.5  &~29.48$\pm$13.96   &~$-$262.32$\pm$28.43~& (7)  &             &     \\ 
                         &                  &                  & 0.2$\pm$4.4       & $-$0.2$\pm$4.4      & (5)  &             &     \\ 
                         &                  &                  & $-$5.9$\pm$12.3   & $-$8.1$\pm$12.2     & (8)  &             &     \\ 
OGLE\,BUL$-$SC32 388121  & 11.1$\pm$----    & 5.0$\pm$----     & $-$78.63$\pm$6.88 &~$-$183.12$\pm$6.66~ & (7)  &             &     \\ 
                         &                  &                  & $-$116$\pm$2      & $-$574$\pm$112      & (1)  &             &     \\ 
                         &                  &                  & $-$114.6$\pm$9.5  & $-$579.6$\pm$9.5    & (5)  &             &     \\ 
                         &                  &                  & 24.9$\pm$7.4      & $-$18.7$\pm$11.6    & (8)  &             &     \\ 
\hline
\end{tabular}
\tablebib{
(1) Monet et al. \citep{2003AJ....125..984M};
(2) Westerlund \citep{1971A&AS....4...51W};
(3) Turon et al. \citep{1992ESASP1136.....T};
(4) Perryman \citep{1997ESASP1200.....P};
(5) R\"oser et al. \citep{2010AJ....139.2440R};
(6) Salim \& Gould \citep{2003ApJ...582.1011S};
(7) Sumi et al. (2004); 
(8) Zacharias et al. \citep{2012yCat.1322....0Z};
(9) van Leeuwen \cite{2007A&A...474..653V}.
}
\end{center}
\end{table*}

\begin{table*}
\caption[]{Details of the IR spectroscopic observations and 
derived spectral types.
Median airmasses are given. The target IDs are the same as in 
Fig.\,\ref{fig:spectra_all}. A+B means that both the primary and 
the secondary were observed, B means that only the secondary was 
observed.}\label{table:spec_obs_log}
\centering
\begin{tabular}{@{}l@{ }c@{ }c@{ }c@{ }c@{ }c@{ }c@{ }l@{ }c@{ }c@{ }c@{ }c@{ }c@{ }c@{}}
\hline\hline
\multicolumn{6}{c}{~~Science Target}                                                                       & ~~~Date~~~~~  & \multicolumn{7}{c}{~~Telluric Standard}                                                         `` \\
Target          & \multicolumn{2}{c}{~~~~Blue grism} & \multicolumn{2}{c}{~~~~Red grism} & Derived\,Sp.\,Types & mm-~~~~~~~& Telluric & \multicolumn{2}{c}{~~~~Blue grism} & \multicolumn{2}{c}{~~~~Red grism} & Sp.  & Ref.  \\
ID              & NDIT$\times$DIT      & sec\,$z$~&\,NDIT$\times$DIT     & sec\,$z$~&~Primary+Secondary~   & ~~~~yyyy  & ID           & NDIT$\times$DIT      & sec\,$z$~&\,NDIT$\times$DIT     & sec\,$z$~     &~Type~&       \\
L\,412-3\,A+B   &  4\,$\times$\,30\,s  & 1.42     &  6\,$\times$\, 30\,s & 1.36     & K2V\,+\,K5V          & 05-2012   & HIP\,084636  &  4\,$\times$\,1.2\,s & 1.46     &  6\,$\times$\,1.2\,s & 1.44          & G3V  & (1)   \\
LHS\,2881\,A+B  & 25\,$\times$\,25\,s  & 1.22     &  7\,$\times$\,100\,s & 1.34     & K3V\,+\,M4V          & 04-2011   & HIP\,064574  &  5\,$\times$\,10\,s  & 1.28     &  5\,$\times$\,10\,s  & 1.39          & G1V  & (1)   \\
LHS\,3188\,B    &  7\,$\times$\,150\,s & 1.10     &  7\,$\times$\,150\,s & 1.05     & n.a.\,+\,M5V         & 04-2011   & HIP\,090446  &  5\,$\times$\,6\,s   & 1.12     &  3\,$\times$\, 6\,s  & 1.07          & G0V  & (2)   \\
LHS\,5333\,A+B  & 12\,$\times$\,2\,s   & 1.14     & 12\,$\times$\,  4\,s & 1.16     & K0V\,+\,M3V          & 05-2012   & HIP\,092515  &  6\,$\times$\,10\,s  & 1.12     &  6\,$\times$\,15\,s  & 1.14          & G2V  & (1)   \\
LTT\,6990\,A+B  &  6\,$\times$\,10\,s  & 1.02     & 12\,$\times$\, 10\,s & 1.02     & K0V\,+\,K4V          & 05-2012   & HIP\,084636  &  5\,$\times$\,2\,s   & 1.04     & 10\,$\times$\,2\,s   & 1.04          & G3V  & (1)   \\
LP\,487-4\,A+B  &  2\,$\times$\,120\,s & 1.01     &  3\,$\times$\,120\,s & 1.01     & G8V\,+\,K5V          & 05-2012   & HIP\,084636  &  5\,$\times$\,2\,s   & 1.04     & 10\,$\times$\,2\,s   & 1.04          & G3V  & (1)   \\
LP\,920-25\,A+B &  6\,$\times$\,30\,s  & 1.39     &  8\,$\times$\, 30\,s & 1.32     & M1.5V\,+\,M1.5V      & 05-2012   & HIP\,084636  &  4\,$\times$\,1.2\,s & 1.46     &  6\,$\times$\,1.2\,s & 1.44          & G3V  & (1)   \\
LP\,922-16\,A+B &  3\,$\times$\,20\,s  & 1.21     &  4\,$\times$\, 30\,s & 1.23     & K1IV\,+\,M1.5V       & 05-2012   & HIP\,092515  &  6\,$\times$\,10\,s  & 1.12     &  6\,$\times$\,15\,s  & 1.14          & G2V  & (1)   \\
M\,124...\,A+B  &  2\,$\times$\,120\,s & 1.01     &  3\,$\times$\,120\,s & 1.01     & M0V\,+\,M1.5V        & 05-2012   & HIP\,098813  & 10\,$\times$\,2\,s   & 1.01     & 20\,$\times$\,2\,s   & 1.01          & G1V  & (3,4) \\
\hline                                                                                                     
\end{tabular}
\tablebib{
(1) Houk \& Smith-Moore \citep{1988mcts.book.....H}; 
(2) Houk \& Cowley \citep{1975mcts.book.....H};
(3) Houk \citep{1982mcts.book.....H};
(4) Gray et al. \citep{2006AJ....132..161G}.
}
\end{table*}

\end{document}